\documentclass[preprint,1p,number,sort&compress]{elsarticle} 

\usepackage{lineno}
\usepackage{amssymb}
\usepackage{amsmath}
\usepackage{eurosym}
\usepackage[normalem]{ulem}
\usepackage{todonotes}

\journal{~}

\begin{document}

%%%%%%%%%%%%%%%%%%%%%%%%%%%%%%%%%%%%%%%%%%%%%%%%%%%%%%%%%%%%%%%%%
%% FRONTMATTER
%%%%%%%%%%%%%%%%%%%%%%%%%%%%%%%%%%%%%%%%%%%%%%%%%%%%%%%%%%%%%%%%%

% title
\title{Optimal combination of energy storages for prospective power supply systems based on Renewable Energy Sources}%
% author information
\author[ne]{Stefan Weitemeyer\fnref{fn1}}%
\author[ne]{David Kleinhans\corref{cor1}}
\ead{david.kleinhans@next-energy.de}%
\author[ne]{Lars Siemer\fnref{fn2}}%
\author[ne]{Carsten Agert}%
\cortext[cor1]{Corresponding author. NEXT ENERGY, EWE Research Centre
  for Energy Technology at the University of Oldenburg,
  Carl-von-Ossietzky-Str.~15, 26129 Oldenburg, Germany. Tel.:~+49 441
  99906 106.}%
\address[ne]{NEXT ENERGY, EWE Research Centre for Energy Technology at
  the University of Oldenburg, 26129 Oldenburg, Germany}
\fntext[fn1]{Current address: Forschungszentrum J\"ulich, 52425 J\"ulich, Germany}
\fntext[fn2]{Current address: Faculty 3 -- Mathematics, University of Bremen, 28359 Bremen, Germany}

% abstract
\begin{abstract}
Prospective power supply systems based on Renewable Energy Sources require measures to balance power generation and load at all times. The utilisation of storage devices and backup power plants is widely suggested for this purpose, whereas the best combination is still to be found. In this work, we present a modelling approach to systematically study scenarios of future power supply systems with a high share of electricity originating from wind and solar resources. By considering backup as a subordinate source of electricity with energy-only costs, the approach is independent of the actual full-load hours of the backup power plants. Applying the approach to multi-year meteorological data for Germany, cost-optimised combinations of storage devices and backup power are identified. We find that even in scenarios with significant excess generation capacities there is a need for storage devices or backup power plants with discharging power on the same order as the average load to balance the system at all times. Furthermore, these capacities seem to be required in some years of the multi-year period only. Our results imply that the societal need of having electricity available at all times can likely be satisfied by installing over-capacities only, whereas a balance has to be found between installing additional backup or storage or generation capacities.
\end{abstract}

% keywords etc (separated by \sep)
\maketitle

%%%%%%%%%%%%%%%%%%%%%%%%%%%%%%%%%%%%%%%%%%%%%%%%%%%%%%%%%%%%%%%%%
%% BODY
%%%%%%%%%%%%%%%%%%%%%%%%%%%%%%%%%%%%%%%%%%%%%%%%%%%%%%%%%%%%%%%%%

%\linenumbers

\section{Introduction} \label{sec:introduction}
Many countries throughout the developed world are currently in the long-term process of changing their electricity supply system from one based on fossil and nuclear fuels to one based on Renewable Energy Sources (RES). Multiple studies have shown that an electricity supply system based entirely on RES is possible -- not only in large-interconnected systems like Europe (e.g.\ \cite{Heide2010, Foundation2010}) and the United States (e.g.\ \cite{Hart2011,Budischak2012, Becker2015}), but also in isolated national systems (e.g.\ \cite{Esteban2012,Mason2010,Connolly2011}).

In many studies dealing with a prospective European power supply systems based on RES, a large share of the electricity will be provided by variable Renewable Energy sources (VRES), which are in particular solar and wind resources \cite{DeVries2007, Jacobson2011}. A power supply system based mainly on VRES requires measures to cope with the natural variability of the power production from these resources in order to ensure our current high security of supply. An option widely considered is the utilisation of storage devices. Here, a combination of small, highly efficient storage capacities for balancing fluctuations in the order of a day or less and large, less efficient seasonal storage capacities for balancing long-term fluctuations appears suitable \cite{Rasmussen2012,Agora2014}.

It is still an open question though, which storage dimensions -- storage capacity and power requirements -- are most suitable for integrating a very high share of VRES into a power supply system. Two large groups of approaches exist to address this question: Many studies rely on cost estimates to come up with possible good combinations of storage and generation capacities (e.g.\ \cite{Budischak2012, Lund2009, Schill2013}). Due to the often relatively high number of parameters and -- generally well-founded -- assumptions, the sensitivity with respect to individual parameters and even more to combinations thereof is often difficult to determine. A different approach is to be as general as possible by considering an abstract view on the system and focusing on the meteorological aspect of future power supply systems, since this aspect will play an important role in future power supply systems based on RES \cite{Bremen2009}. Using mainly long-term RES power production time-series and corresponding load data, these works have investigated e.g.\ the large-scale need for the installation of storage capacities in a 100\%-RES scenario for Europe and how these needs can be reduced using excess generation as well as transmission capacities \cite{Heide2010,Heide2011,Rasmussen2012,Rodriguez2014}.

Besides the energy capacity of the storage devices, their power requirements are also an important dimension to be considered for future power systems based on (V)RES. Weiss et al.\ investigated the storage requirements for the German power supply system in which 80\% of the total electricity consumption would be provided by a combination of solar, wind and hydropower resources; and it was found that in this case 31 GW of discharging power and slightly less for charging power would be required for the long-term storage devices when combined with the current pumped-hydro capacity of 8 GW as short-term storage \cite{Weiss2013a}. Like the need for the energy capacity of the storage devices rises with increasing share of RES penetration, so does the need for storage power. Using a holistic model with a hierarchical management approach, Bussar et al.\ recently found that a renewable power supply system for the EU-MENA region (Europe, Middle East and North Africa) would be most economic with a combination of storage devices having a total discharging power on the same order as the peak load, whereupon long-term storage was also found to need higher charging power than discharging power \cite{Bussar2016}. Furthermore, the discharge power of the long-term storage was found to be higher than the combined discharge power of both short-term and medium-term storage \cite{Bussar2016}, which stresses the importance of long-term storage in future power systems based on (V)RES. 

In this work, we present an optimisation approach to systematically study power systems with multiple storage systems, well defined by their energy content, charging and discharging power as well as their efficiency. Here, a special focus is set on the power requirements of the storage systems. The modelling approach is an advancement of our previous modelling approaches (cf.\ \cite{Weitemeyer2014, Weitemeyer2016}) and is intended to provide deeper insights into urgent questions in the design of prospective power systems based on energy generated from fluctuating renewable sources.

%%%%%%%%%%%%%%%%%%%%%%%%%%%%%%%%%%%%%%%%%%%%%%%%%%%%%%%%%%%%%%%%%
%% MODEL
%%%%%%%%%%%%%%%%%%%%%%%%%%%%%%%%%%%%%%%%%%%%%%%%%%%%%%%%%%%%%%%%%

\section{Modelling multiple storage systems}\label{sec:modelling}
To investigate combinations of energy storages for the integration of high shares of VRES into power systems, we suggest a novel modelling approach using an economic optimisation procedure. For our approach we build on previous contributions on the integration of VRES in Germany and Europe \cite{Weitemeyer2014,Weitemeyer2016}. While these contributions were restricted to meteorological aspects of the integration and were by construction limited to the investigation of the effects of one single storage class, the optimisation approach developed here will allow to study economic aspects and power systems with multiple storages in a yet rather straightforward and instructive manner.

\subsection{Optimisation of storages and backup}
\label{ssec:Optimisation}
We start from time series data of the residual load for a certain area of interest, $R(t_i)$, sampled at time lag $\Delta t$, i.e.\ $t_i=t_1 +\Delta t (i-1)$ with $i \in \{1,2,\ldots,N\}$. $R(t_i)$ can be obtained by subtracting the power generation of both must-run units and variable, non-dispatchable renewable energy sources from the load, details will be described later in section \ref{ssec:resources}. If $R(t_i)<0$ there is a surplus of energy which could be curtailed, whereas at times $t_i$ with $R(t_i)>0$ there is a lack of energy which needs to be compensated. For balancing of the loads -- in particular in the second case -- it is aimed to use either backup power plants or storages, with the latter being charged in times of surplus of energy. 

A storage $j$ is characterised by its charging ($P_j^{CL}$) and discharging ($P_j^{DL}$) power, its round-trip efficiency\footnote{The round-trip efficiency $\eta$ is split equally between charging and discharging in this work. Individual values for charging efficiency ($\eta_j^c$) and discharging efficiency ($\eta_j^d$) can easily be implemented but would only influence the interpretation of the value of the storage capacity $H_j^{\max}$. The storage capacity $H_j^{\max}$ corresponds to the energy stored, e.g.\ in the form of chemical energy, while $\eta_j^c$ and  $\eta_j^d$ determine the flow of electrical energy from/to the overall system (cf.~eq.~(\ref{eq:storagelevel})).} $\eta_j$, and its maximum capacity in terms of energy $H_j^{\max}$. We assume that each of these parameters implies costs, which are characterised by their respective equivalent annual installation and operating costs per unit, $q_j^{CL}$, $q_j^{DL}$, and $q_j^{H}$. If we for the time being neglect grid limitations and grid losses, storage units with the same technology (i.e.\ in particular with the same efficiencies for charging and discharging) can be combined to a single storage, which significantly reduces the amount of different storages which need to be considered.

The approach we would like to propose can be characterised as an economic optimisation approach to identify which combination of storages and backup power would be most beneficial from an economic perspective (i.e.\ ''cheapest``) to guarantee a non-positive residual load at each moment in time. The costs for installing and operating the generation capacities are explicitly not included in the optimisation approach, since these costs depend on the scenario considered and be can calculated and added separately (c.f.~section~\ref{ssec:resources}). Other costs of prospective power systems which are not included in our approach (e.g.\ transmission costs, social costs) need to be investigated and compared in more detailed analyses in future work.

The linear optimisation problem with the corresponding cost function $Q_{SB}$ reflecting the annual costs for storage and backup is defined as:

\begin{align}
  \min\limits_{\substack{P_j^{CL},\, P_j^{DL},\\H_j^{0},\, H_j^{\max},\\ P_j^{c}(t_i),\, P_j^{d}(t_i),\\ B(t_i) \,~\forall j,\\i=1,\dots ,N}} & Q_{SB}=P_j^{CL} \cdot q_j^{CL}+P_j^{DL}\cdot q_j^{DL}+H_j^{\max}\cdot q_j^{H} + \frac{\Delta t}{N_a}\cdot \sum_{i=1}^{N}B(t_i)\cdot q^B \tag{OP} \label{eq:costfunction}\\
\mbox{subject to:} \quad & R_{\alpha,\gamma}\left(t_i\right) + \sum_{j=1}^{m}{\left( P_j^{c}(t_i)-P_j^{d}(t_i)\right)}-B(t_i)\leq 0\,,\forall t_i \label{eq:cond_resload} \\
& 0 \le P_j^{c}(t_i)\le P_j^{CL}\,, \ \forall t_i, j=1, \dots ,m\\
& 0 \le P_j^{d}(t_i)\le P_j^{DL}\,, \ \forall t_i, j=1,\dots,m\\
& 0 \le S_{j} \left(t_i \right)\le H_j^{\max}\,, \ \forall t_i,j=1,\dots,m \\
& 0 \le H_j^{0}\leq S_{j} \left(t_i=N \right)\,, \ j=1,\dots,m\\
& 0 \le B(t_i)\le P_B^{DL}\,, \ \forall t_i
\label{eq:limitbackup}
\end{align}
Here, the -- generally not unique -- time series $ P_j^{c}(t_i)$, $P_j^{d}(t_i)$ and $B(t_i)$ determine the charge and discharge of the respective storage and the backup power required (i.e. the operation of storages and backup). For the backup, $B(t_i)$ describes the actual time series of backup usage, $N_a:=N\Delta t/\mathrm{a}$ is the length of the time series in multiples of years, and $q^B$ are the costs for backup per energy unit. The backup energy is modelled in this rather simple way to be able to include a wide variety of backup options, ranging from biogas power plants with high usage rates to peak power plants with low number of full-load hours as well as load shedding options, in a rather straightforward manner. A sufficiently high value for the backup costs $q^B$ implies that backup energy is not primarily used in comparison to wind and solar energy. The real costs for some of the technologies considered as backup power might be cheaper if they achieve a high number of full-load hours. However, the latter is not known until the optimisation process is finished. For most of the investigated scenarios in this work, the maximum backup power $P_B^{DL}$ is not limited so as to include the wide variety of backup options mentioned above. A closer look at the backup power plants is taken in section \ref{ssec:minimal_backup}.

The constraints \eqref{eq:cond_resload} -- \eqref{eq:limitbackup} guarantee that the parameters and the storage level
\[S_{j} \left(t_i\right):= H_j^{0}+\Delta t\cdot\left(\sqrt{\eta_j}\sum_{k=1}^{i}P_j^{c}(t_k)-\frac{1}{\sqrt{\eta_j}}\sum_{k=1}^{i}P_j^{d}(t_k)\right) \tag{7}\label{eq:storagelevel}\]
remain within the boundaries associated with the minimum required installation costs, namely (1) meeting the residual load at all times by (net) discharging the storages and utilisation of backup power\footnote{Curtailment of the VRES generation is also allowed as the residual load may become negative.}, the (2) charging and (3) discharging power must not exceed the charging and discharging power limit of the respective storage, (4) the storage level for each storage is always between zero (empty) and the size of the storage $H_j^{\max}$, (5) the storage levels at the end of the simulation must not be lower than at the beginning, and (6) backup power plants can only provide positive backup power. The structure of \eqref{eq:costfunction} is related to a linear minimisation problem also referred to as linear program. The main difference to a deterministic linear optimisation problem is the fact that the upper boundaries of the constraints are not fixed during the optimisation process. To solve \eqref{eq:costfunction} with flexible constraints, standard solver for linear programs (e.g.~simplex or interior point methods) can be applied on piecewise equivalent subproblems \cite{nocedal2006numerical, grossmann1983effective}.

Additionally, the approach allows integration of other power plants by integrating them as a (virtual) storage. The storage size $H_j^{\max}$ corresponds to the highest possible energy production in the period of investigation, while the charging power $P_j^{CL}$ of the corresponding storage / power plant would be set to zero, that is $P_j^{CL}=0.$ This procedure is not applied in this work. 

Overall, the optimisation approach can be seen as an investment decision approach with a perfect and unlimited prediction horizon, while the actual dispatch of the individual units would have to be decided upon in a more detailed simulation with a much smaller prediction horizon.

\subsection{\label{ssec:resources}Meteorological time series, residual loads and total system costs}
For prospective power systems the residual load depends strongly on the amount of non-dispatchable solar and wind energy in the system. Provided that time series for the load ($L(t_i)$) and for the availability of solar ($S(t_i)$) and wind ($W(t_i)$) resources are available, it is straightforward to investigate the implications of their installation on the residual loads. For this purpose we follow a procedure outlined in \cite{Weitemeyer2014}.

With $\langle x\rangle_t:=N^{-1}\sum_{i=1}^Nx(t_i)$  representing the long-term average of a time series, the residual load $R_{\alpha,\gamma}(t_i)$ is modelled by

\[R_{\alpha,\gamma}(t_i)=L(t_i)-\gamma \cdot \langle L\rangle_t \left(\alpha \frac{W(t_i)}{\langle W\rangle_t}+\left ( 1 - \alpha \right ) \frac{S(t_i)}{\langle S\rangle_t}\right ) \]

Here, the respective shares of wind and solar power generation of the gross electricity demand are given by $\gamma\alpha$ and $\gamma(1-\alpha)$. The parameter $\gamma$ is termed the average renewable energy power generation factor and determines the total electricity which can be produced from VRES (cf.~\cite{Heide2011}). A value of $\gamma=1.0$ corresponds to a scenario, in which the long-term averaged total generation from solar and wind resources is equal to the average demand.

In this work, we concentrate on the technical aspect of a future power system. Hence, we focus in our model on the costs for the installation and the operation of combinations of generation, storage and backup capacities. For every investigated scenario, the total equivalent annual costs $Q_{tot}$ are given by the costs for storage and backup $Q_{SB}$ -- as reflected by the cost function of \eqref{eq:costfunction} -- and the costs for the generation capacities $Q_G$

\[ Q_{tot} = Q_{SB} + Q_{G} \]

The generation costs $Q_G$ are given by 

\[ Q_G= P_W \cdot q^W + P_S \cdot q^S + Q_{MR}\quad.\]

Here, the parameters $P_W$ and $P_S$ describe the respective installed capacities of wind and solar as given by the wind share $\alpha$ and the generation factory $\gamma$ for each scenario. The parameters $q^W$ and $q^S$ describe the equivalent annual installation and operating costs per power unit, which are fixed for each scenario (cf.\ section \ref{ssec:data}). The parameter $Q_{MR}$ reflects the costs for must-run units; the latter are not considered in this work. Overall, the generation costs are fixed for each investigated scenario and not relevant for the optimisation process described in the section \ref{ssec:Optimisation}.

%%%%%%%%%%%%%%%%%%%%%%%%%%%%%%%%%%%%%%%%%%%%%%%%%%%%%%%%%%%%%%%%%
%% APPLY TO GERMANY
%%%%%%%%%%%%%%%%%%%%%%%%%%%%%%%%%%%%%%%%%%%%%%%%%%%%%%%%%%%%%%%%%

\section{Storage requirements for Germany}\label{sec:germany-base}

The modelling approach described in section \ref{sec:modelling} is now applied to data for Germany. Germany is chosen in this work as it has significantly increased its RES share over the last years with more than 32\% of electricity originating from RES in 2015 \cite{BMWi2016}. Furthermore, it has also been the subject of our previous publication dealing with a more meteorologically focused perspective on the storage demand in future renewable power systems \cite{Weitemeyer2014}.

\subsection{Data description}\label{ssec:data}
The modelling approach requires long-term power generation for renewable energy sources as well as corresponding load data. In this work, we use wind and solar power generation data with hourly resolution ($\Delta t$=1h). The division between wind and solar power generation is determined by the wind share $\alpha$ (cf.\ section \ref{ssec:resources}), while for the wind data itself holds that 52\% of the installed capacity is onshore. The data sets span the year 2006-2012 ($N_a$=7) and thus contain N=61368 data points each. This way, the variability of the power production from VRES over multiple years is included in the data. The load data originate from the data from the transmission system operators (TSOs) in Germany for the same period and are available from ENTSO-E. The average load in the period of investigation is $\langle L\rangle_t\approx55.1~\text{GW}$, subsequent energy values are normalised where appropriate to average load hours, abbreviated av.l.h., with 1 av.l.h.~$\approx$ 55.1 GWh. The data sets were also used in previous publications \cite{Weitemeyer2014,Kies2015}. More details regarding the wind and solar power generation data sets can be found in \cite{Kies2015}.

Three different storage technologies are used in this work to balance fluctuations. First are pumped hydro storages, which are nowadays mainly used in Germany to balance fluctuations on the scale of multiple hours (PHS, $\eta_{PHS}=82 \%$ \cite{Adamek2012}). The current installed capacity is around 39 GWh according to \cite{Eurelectric2011} with a small potential for additional capacities \cite{Steffen2012}. Given Germany's proximity to the Alps region and the large hydro facilities there, we assume a maximum pumped-hydro storage size available to the German power system as $H_{PHS}^{\max}$ = 4 av.l.h.\ $\approx$ 220.6 GWh. As this value is still below the required storage capacities found in other works investigating fully renewable power supply systems(\cite{Heide2010, Heide2011, Weitemeyer2014}), further storage technologies are considered. Batteries based on lithium-ion (LIB, $\eta_{LIB}=88 \%$ \cite{Adamek2012}) are the second storage technology considered in our work. Finally, as a rather long-term storage we use synthetic hydrogen stored in caverns (H2S, $\eta_{H2S}=45 \%$ \cite{Adamek2012}). By not including any restrictions on the operation of the storage devices, all storages can in principle be used as short-term and long-term storage.

The equivalent annual costs $q^X$ for the generation and storage units is composed of installation costs as well as annual operating and maintenance (O\&M) costs. For consistency with respect to the underlying assumptions, the data are to a large extend based on a single source and are assumed costs for Germany by mid-century. More precisely, the costs for storages and generation capacities used in this work are based on \cite{Adamek2012} (in turn based on \cite{Nitsch2010}) as well as the updated version \cite{Nitsch2012c} and shown in table \ref{tab:parameters}. The prices $q^X$ are achieved by converting these costs to equivalent annual costs using a fixed interest rate of r=6\% throughout the work. Details regarding this conversion to equivalent annual costs can be found in \ref{sec:app_eac}. 

The costs for backup energy is generally set to $q^B$=0.15 EUR / kWh\footnote{This value corresponds e.g.\ to costs for backup power originating from CCGT in \cite{Nitsch2010} and costs for biomass power plants in \cite{Kost2013}.}. As mentioned already above in section \ref{ssec:Optimisation}, the backup energy is modelled this way to include a wide variety of backup options.

\begin{table}[htbp]
  \centering
  \caption{Parameters for base case of economic evaluation: $I^{X}$ are installation costs, $f^{O,X}$ describe O\&M costs in relation to installation costs, n is the lifetime (cf.\ \ref{sec:app_eac}), data are based on \cite{Adamek2012, Nitsch2010, Nitsch2012c}.}
  \label{tab:parameters}
  \begin{tabular}{lrrcc}
    \hline
    Technology & \multicolumn{2}{c}{$I^{X}$} & $f^{O,X}$&  n \\
     & \euro~/ kW & \euro~/ kWh & \% & years \\
    \hline
    wind (onshore) & 1000 & & 4.0 & 18 \\
    wind (offshore) & 1400 & & 5.5 & 18 \\
    photovoltaic & 850 & & 1.0 & 20 \\
    \hline
    PHS (capacity) & & 10  & 1.0 & 80 \\
    PHS (charging) & 250 & & 1.0 & 35 \\
    PHS (discharging) &250 & & 1.0 & 35 \\
    LIB (capacity) & &150& 0.5 & 25 \\
    LIB (charging) &25& & 0.5 & 30 \\
    LIB (discharging) &25 & & 0.5 & 30 \\
    H2S (capacity) & &0.7& 2.5& 40 \\
    H2S (charging) &350& & 2.5& 25 \\
    H2S (discharging) &700& & 2.5& 25 \\ 
\end{tabular}
\end{table}

\subsection{Total system costs}\label{ssec:generalalphagamma}

Let us start our investigation of the cost-optimal design of the prospective German power supply system by studying the influence of the wind share $\alpha$ and the power generation factor $\gamma$ on the total system costs $Q_{tot}$. For this, we use the cost function as defined in equation \eqref{eq:costfunction} and consider the three storage technologies mentioned in section \ref{ssec:data}.

\begin{figure}[htb]
  \centering
    \includegraphics[width=0.95\textwidth]{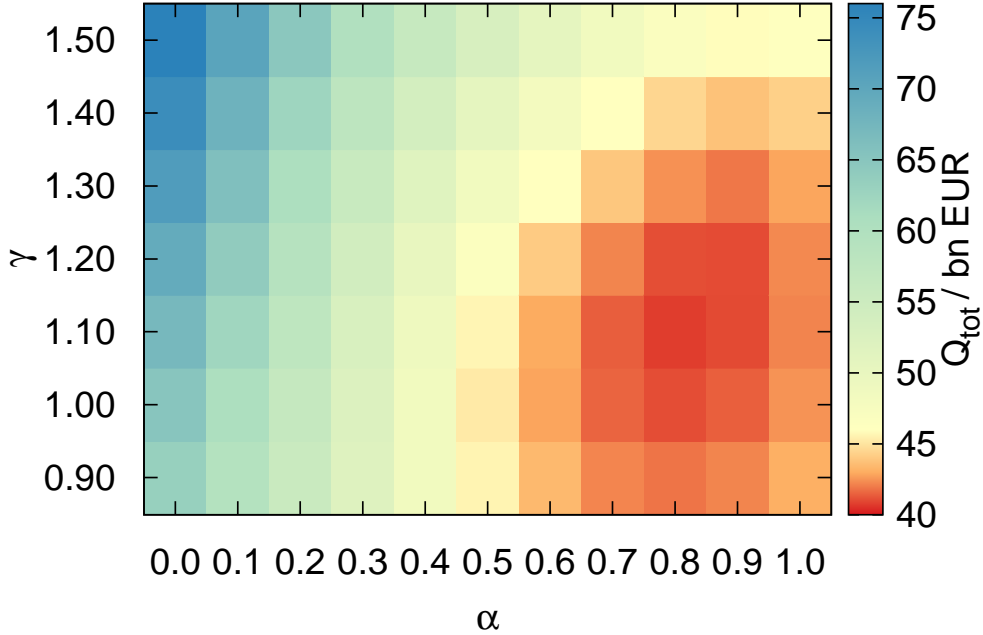}
  \caption{Heatmap of equivalent annual costs $Q_{tot}$ for scenarios with different values of wind share $\alpha$ and generation factor $\gamma$. For all scenarios, the pumped-hydro storage size was limited to $H_{PHS}^{\max}$ = 4 av.l.h.\ $\approx$ 220.6 GWh and the backup power was in practical terms unlimited ($P_B^{DL}=10\cdot\langle L\rangle_t$).}
    \label{fig:heatmap_costs}
\end{figure}

Figure \ref{fig:heatmap_costs} shows the total equivalent annual costs $Q_{tot}$ for scenarios with different combinations of $\alpha$ and $\gamma$ considering all three storage technologies ($j=3$). In these scenarios, the size of the pumped-hydro storages was limited to $H_{PHS}^{\max}$ = 4 av.l.h.\ $\approx$ 220.6 GWh\footnote{This limitation corresponds to an additional constraint for \eqref{eq:costfunction}: $H_{PHS}^{\max}\leq \mathrm{4~av.l.h.}$} in order to reflect the limited resources for pumped hydro (see also section \ref{ssec:data}) and the backup power was in practical terms unlimited ($P_B^{DL}=10\cdot\langle L\rangle_t$). As we can see from the figure, the sensitivity with respect to the wind share $\alpha$ is higher than the sensitivity with respect to the capacity factor $\gamma$. As an example, for a fixed generation factor $\gamma=1.0$, the total system costs $Q_{tot}$ vary from $Q_{tot}\approx$~65.1 bn EUR as equivalent annual costs for a solar-only scenario ($\alpha$=0.0) to $Q_{tot}\approx$~41.1 bn EUR as equivalent annual costs for a scenario with $\alpha=0.8$. With respect to the generation factor $\gamma$, the costs variation is significantly less, e.g.\ for $\alpha=0.8$ from $Q_{tot}\approx$~40.8 bn EUR as equivalent annual costs for $\gamma=1.1$ to $Q_{tot}\approx$~46.9 bn EUR as equivalent annual costs for $\gamma=1.5$. 

Overall, this figure illustrates that the total equivalent annual system costs $Q_{tot}$ as defined in this work are strongly influenced by the costs for different technology (cf.\ table~\ref{tab:parameters}).

Based on the previous results, we will now investigate scenarios with two different wind shares as well as different capacity factors and draw comparisons between them. For this, we will focus on scenarios with a wind share of $\alpha=0.80$ (cheapest option and also share in related works (e.g.\ \cite{Kies2015}) as well as scenarios with a high solar share, that is $\alpha=0.50$, in the upcoming investigations.

\subsection{Optimal storage combinations}\label{ssec:basecase}
\begin{figure}[htb]
  \centering
    \includegraphics[width=0.95\textwidth]{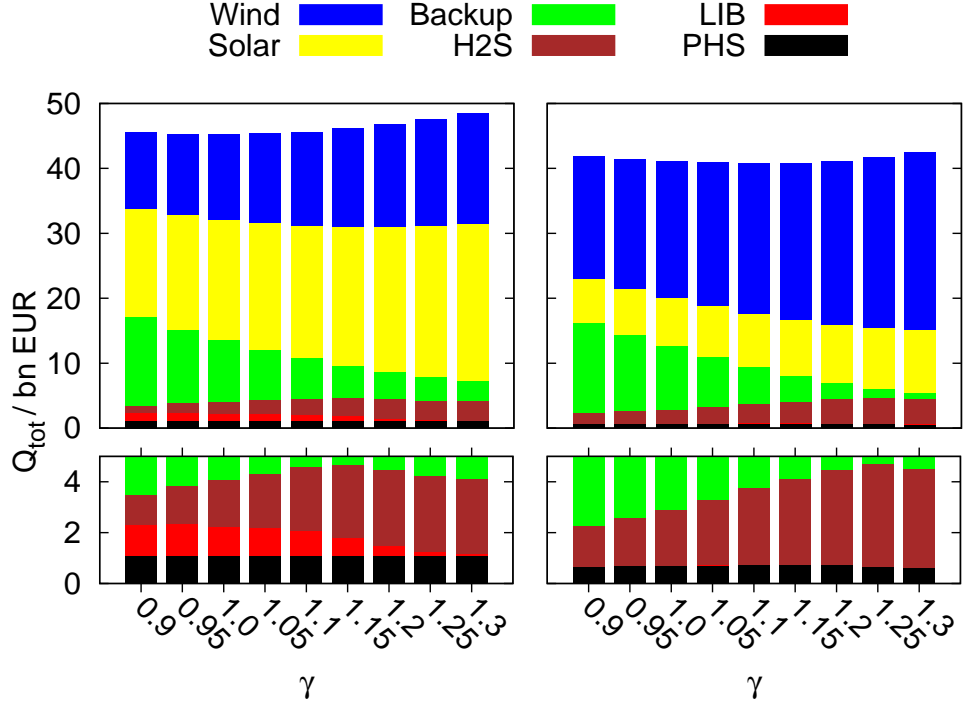}
  \caption{Total equivalent annual costs $Q_{tot}$ and its breakdown on system components for different values of generation factor $\gamma$ and a wind share $\alpha=0.50$ (left panel) and a wind share $\alpha=0.80$ (right panel). The lower panels show a detail of the upper panels for a more straightforward discussion of the costs originating from storages. For all scenarios, the pumped-hydro storage size was limited to $H_{PHS}^{\max}$ = 4 av.l.h.\ $\approx$ 220.6 GWh and the backup power was in practical terms unlimited ($P_B^{DL}=10\cdot\langle L\rangle_t$).}
    \label{fig:histogram_cost_basecase}
\end{figure}

The total systems costs illustrated in figure \ref{fig:heatmap_costs} are made up of different components. Figure \ref{fig:histogram_cost_basecase} shows the total equivalent annual costs $Q_{tot}$ and its breakdown to system components for different values of the generation factor $\gamma$ as well as for a wind share $\alpha=0.50$ (fig.\ \ref{fig:histogram_cost_basecase}, left panel) and a wind share $\alpha=0.80$ (fig.\ \ref{fig:histogram_cost_basecase}, right panel). It can be seen in this figure that the share of costs related to storage capacities in all investigated scenarios make up to only 11\% of the total costs, while in all scenarios the highest share of costs is related to wind and solar generation capacities as well as backup costs. The share of costs related to generation and backup is higher than the results found by Bussar et al., where those costs were responsible for about 74\% of the total costs (excluding grid costs). This is likely due to differences in the particular costs assumptions. The overall trend that the major shore of costs is related to the production costs is consistent in both works though.

Furthermore, the figure illustrates that increasing the generation factor $\gamma$ significantly decreases the need for energy originating from backup power plants, while the costs for storage and hence its capacities are increased. As already seen in figure \ref{fig:heatmap_costs}, the overall costs $Q_{tot}$ increase slightly only when the generation factor $\gamma$ is changed, implying that the increased costs for generation capacities are approximately equivalent to the combined costs of decreased costs for backup and increased costs for storage capacities.

Besides the overall difference in costs related to the higher generation costs in the high-solar scenario ($\alpha=0.50$), a major difference between the scenarios with different wind shares $\alpha$ is the presence of lithium-ion batteries in scenarios with a wind share of $\alpha$=0.50, while no lithium-ion batteries at all are present in the results for scenarios with a wind share of $\alpha=0.80$. This can be explained by considering that lithium-ion batteries have rather low charging and discharging power costs and high capacity costs, and these are favourable for scenarios with a relatively high solar-share and hence strong daily fluctuations. On the contrary, the cost-minimised setup in wind-dominated scenarios requires the utilisation of hydrogen storage which is characterised by comparably high charging and discharging power costs and low capacity costs.

The aforementioned characteristics of the different storage technologies can also be identified when further splitting the storage costs into costs for charging and discharging power on one side and capacity on the other side. For all the scenarios shown in figure \ref{fig:histogram_cost_basecase} the equivalent annual costs for charging and discharging power are significantly higher than the respective equivalent annual costs for capacity when pumped-hydro storage or synthetic hydrogen storage is considered. On the contrary, the equivalent annual costs for storage capacity of the lithium-ion batteries are found to be up to ten times the equivalent annual costs for the charging and discharging power.

\begin{figure}[htb]
  \centering
    \includegraphics[width=0.95\textwidth]{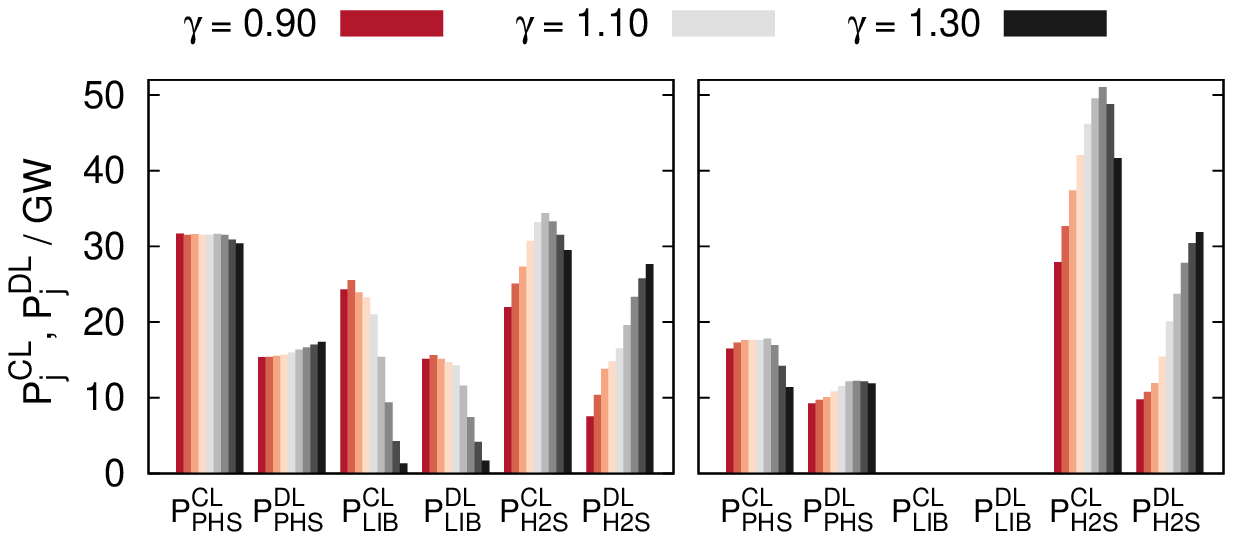}
  \caption{Charging $P_j^{CL}$ and discharging power $P_j^{DL}$ for pumped-hydro, lithium-ion and hydrogen storage for different values of generation factor $\gamma$ and a wind share $\alpha=0.50$ (left panel) and a wind share $\alpha=0.80$ (right panel). In each block, the generation factor $\gamma$ increases from $\gamma$=0.90 to $\gamma$=1.30 in step sizes of 0.05. For all scenarios, the pumped-hydro storage size was limited to $H_{PHS}^{\max}$ = 4 av.l.h.\ $\approx$ 220.6 GWh and the backup power was in practical terms unlimited ($P_B^{DL}=10\cdot\langle L\rangle_t$).}
    \label{fig:PcPd_basecase}
\end{figure}

With the actual power dimensioning of the storage being a focus of this work, figure \ref{fig:PcPd_basecase} examines the previous results in more detail and illustrates the corresponding $P_j^{CL}$ and discharging power $P_j^{DL}$ for pumped-hydro storage, lithium-ion storage and hydrogen storage. A wind share of $\alpha=0.50$ was used for the left panel and a wind share of $\alpha=0.80$ for the right panel.

For scenarios with a wind share $\alpha=0.50$, the charging $P_{PHS}^{CL}$ and discharging power $P_{PHS}^{DL}$ of the pumped-hydro storage is almost irrespective of the generation factor $\gamma$, whereas the discharging power $P_{PHS}^{DL}$ is always lower than the corresponding charging power $P_{PHS}^{CL}$. This can be understood by considering that there are clear daily patterns in the residual load of these high-solar scenarios with comparably high surpluses during the day. The charging $P_{LIB}^{CL}$ and discharging power $P_{LIB}^{DL}$ for the lithium-ion batteries reach their respective maximum value for a generation factor of $\gamma=0.95$ and decreases for larger values of $\gamma$. For the latter, the deficit hours in the residual load are reduced, furthermore the peaks of the surplus generation at midday can be curtailed, leading to a reduced need for the rather expensive lithium-ion batteries. For scenarios with a generation factor $\gamma \geq 1.0$, the charging $P_{H2S}^{CL}$ and discharging power $P_{H2S}^{DL}$ of the hydrogen storage exceeds the corresponding values of the lithium-ion batteries. This implies that the hydrogen storage replaces the lithium-ion storage in some respects. Furthermore, it is worth mentioning that the combined discharging power of all three storages increases with increasing generation factor $\gamma$, reaching a combined sum almost as high as the average load $\langle L\rangle_t$.

Comparing the high-solar scenario ($\alpha=0.5$, figure \ref{fig:PcPd_basecase}, left panel) with the wind-dominated scenario ($\alpha=0.8$, right panel) shows that the charging $P_{PHS}^{CL}$ and discharging power $P_{PHS}^{DL}$ of the pumped-hydro storage is significantly lower in the scenario with $\alpha=0.8$. Furthermore, the synthetic hydrogen is found to be the storage technology with the highest charging $P_{H2S}^{CL}$ and discharging power $P_{H2S}^{DL}$.

The results presented in this section are meant to represent a base case of our investigations. Based on these results, we will continue by studying three different aspects of the scenarios. First, the backup energy and its power requirements (section \ref{ssec:minimal_backup}). Second, the need of seasonal storage (section \ref{ssec:noptg}) as well as, third, the influence of the yearly variations on the results (section \ref{ssec:yearlyvariations}).

\subsection{Minimal backup power}\label{ssec:minimal_backup}
In the previous investigations in sections \ref{ssec:generalalphagamma} and \ref{ssec:basecase}, the backup power $P_B^{DL}$ was not limited so as to include a variety of backup technologies. That was also the reason why we chose an energy-only price for backup usage and did not include any power price. In this section, we aim to limit the backup power and investigate the resulting effects on the average equivalent full-load hours of the backup power plants (cf.\ condition \ref{eq:limitbackup} in section \ref{sec:modelling}). For this, the average full-load hours of the backup power plants are defined as

\[
\text{FLH}=\frac{\langle B \rangle_t\cdot 8760}{P_B^{DL}}\,.\] 

\begin{figure}[htb]
    \centering
    \includegraphics[width=.95\textwidth]{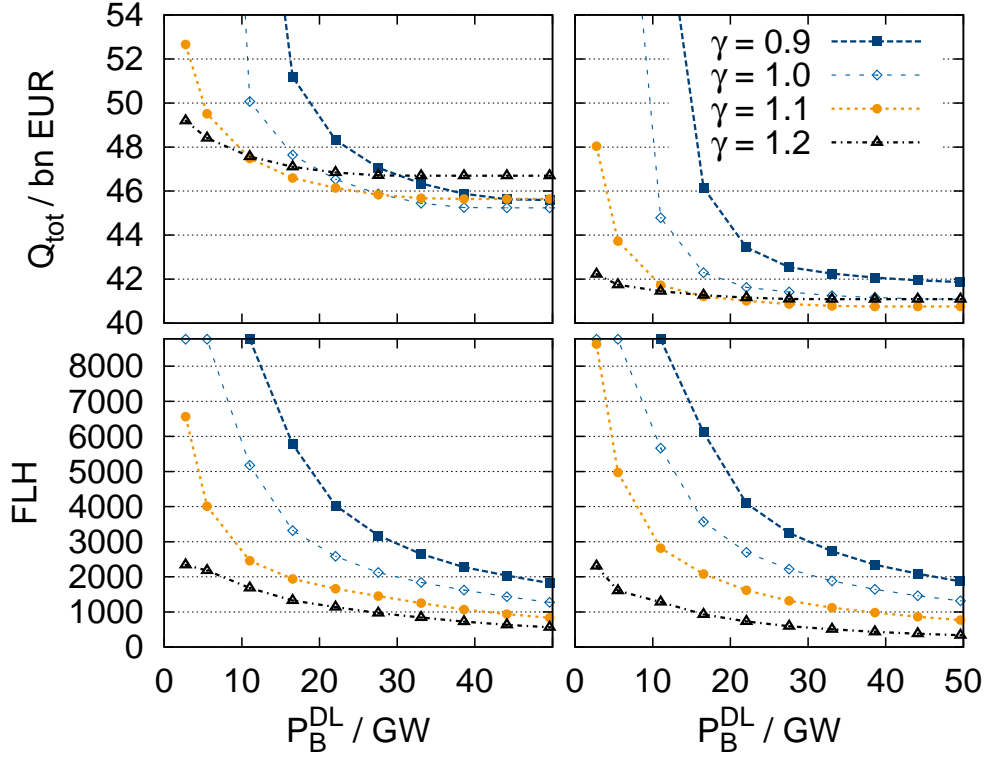}
  \caption{Total equivalent annual costs $Q_{tot}$ (top) and corresponding average full-load hours per year FLH (bottom) as function of backup power limit $P_B^{DL}$ for different values of generation factor $\gamma$: Wind share is $\alpha=0.50$ (left panel) and $\alpha=0.80$ (right panel). For all scenarios, the pumped-hydro storage size was limited to $H_{PHS}^{\max}$ = 4 av.l.h.\ $\approx$ 220.6 GWh.}
\label{fig:backup_fullload}
\end{figure}

Figure \ref{fig:backup_fullload} shows the total equivalent annual costs $Q_{tot}$ (top) and corresponding average full-load hours FLH (bottom) as a function of the backup power limit $P_B^{DL}$ for different values of the generation factor $\gamma$, where a wind share $\alpha=0.5$ was used for the left and a wind share $\alpha=0.8$ was used for the right panel.

Starting with the upper end of the investigated backup power limits (right side of each half, $P_B^{DL}\approx$ 50 GW), the total equivalent annual costs $Q_{tot}$ is almost equal to the corresponding costs without backup power limits. In this case, the average full-load hours reach values of up to about 1890 equivalent full-load hours per year for $\gamma=0.9$ and $\alpha=0.8$, slightly less for $\alpha=0.5$, and as low as about 335 equivalent full-load hours for $\gamma=1.2$ and $\alpha=0.8$. Decreasing the backup power limit down to $P_B^{DL}\approx$ 30 GW only increases the total equivalent annual costs $Q_{tot}$ for all investigated  scenarios. However, even then the backup power plants reach about 3000 equivalent full-load hours only. Reduction of the backup power limit $P_B^{DL}$ below this threshold leads to a significant increase in total equivalent annual costs $Q_{tot}$. The most extreme values of $Q_{tot}$ should be ignored, as the backup power plants are almost used permanently in these scenarios, which would presumably result in production costs much below the fixed costs used in the approach and therefore does seem to be inconsistent with the assumptions.

In summary, the investigations in this section show that a compromise between the installation of excess generation capacities and backup power plant capacities is required. 

\subsection{Systems without hydrogen storage}\label{ssec:noptg}

For our base case scenarios, we assumed that the synthetic hydrogen storage technology will make large advancements in comparison to today and that cheap storage capacities in underground caverns will become available, leading to the costs assumptions in table \ref{tab:parameters}. Using these assumptions, the hydrogen storage was found to be installed in large capacities, in particular in the scenarios with a wind share $\alpha=0.8$ (cf.\ section~\ref{ssec:basecase}). To further study the role of hydrogen storage, we also investigated scenarios where the proposed hydrogen technology does not exist, that is only pumped-hydro storage and lithium-ion batteries exist as storage technologies. Using this restriction, we find that the total equivalent annual costs $Q_{tot}$ increase significantly in most of the investigated scenarios. As an example, for scenarios with a wind share of $\alpha=0.5$ the total equivalent annual costs $Q_{tot}$ increase from $Q_{tot}\approx$~45.2 bn EUR to $Q_{tot}\approx$~46.8 bn EUR for $\gamma$=1.0 , and from $Q_{tot}\approx$~48.5 bn EUR to $Q_{tot}\approx$~51.4 bn EUR for $\gamma$=1.3. The increase in costs is largely due to an increase in costs for lithium-ion batteries as well as an increased backup usage. Reducing the backup costs to $q^B$=0.12 EUR / kWh to take into account the latter effect was found to reduce the total equivalent annual costs $Q_{tot}$, e.g.\ for a wind share $\alpha=0.50$ and a generation factor $\gamma$=1.0 to $Q_{tot}\approx$~44.4 bn EUR. 

For scenarios with a wind share $\alpha=0.8$, the increase in costs for scenarios without hydrogen storage is even higher. As an example, the total equivalent annual costs $Q_{tot}$ increase from $Q_{tot}\approx$~41.1 bn EUR to $Q_{tot}\approx$~43.9 bn EUR for $\gamma$=1.0 , and from $Q_{tot}\approx$~42.4 bn EUR to $Q_{tot}\approx$~47.2 bn EUR for $\gamma$=1.3. In these cases, the lack of hydrogen storage is compensated to a large extent by increasing usage of backup power, while also the capacities for pumped-hydro storage are increased and small capacities of lithium-ion batteries are installed. In this case, the reduction of the backup costs to $q^B$=0.12 EUR / kWh led to a decrease in total system costs for e.g.\ a generation factor $\gamma$=1.0 to $Q_{tot}\approx$~40.1 bn EUR. 

An iterative process could be used to adjust the backup costs to the resulting full-load hours for different backup technologies. However this would contradict the simple design of our approach and is beyond the scope of this work.

\subsection{Influence of length of investigation period}\label{ssec:yearlyvariations}

As a final variation of our base case scenarios, we investigate an aspect which is related to the meteorological perspective of the investigated scenarios. In contrast to the previous parts of this work, we treat each year of the scenarios individually in this section, so as to reduce the statistical variation of the original data set. For this purpose, all years are treated individually, implying that the variations in average load are eliminated by normalisation and the installed capacities for wind and solar generation capacities are adjusted to the respective generation pattern in the corresponding year.

\begin{figure}[htb]
  \centering
    \includegraphics[width=0.95\textwidth]{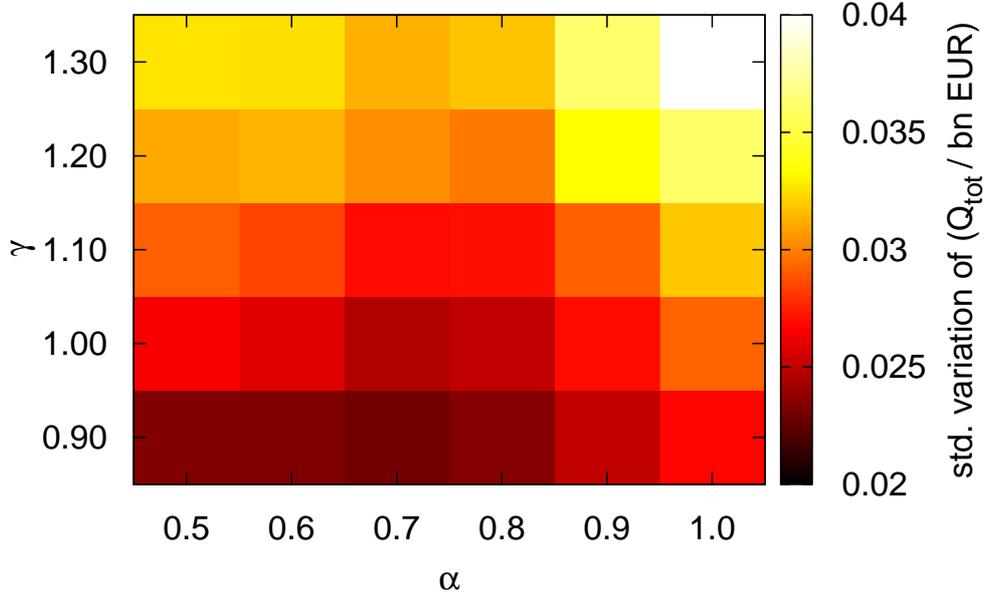}
  \caption{Standard variation of the total equivalent annual system costs $Q_{tot}$ between different years for different values of the wind share $\alpha$ and the generation factor $\gamma$. For all scenarios, the pumped-hydro storage size was limited to $H_{PHS}^{\max}$ = 4 av.l.h.\ $\approx$ 220.6 GWh and the backup power was in practical terms unlimited ($P_B^{DL}=10\cdot\langle L\rangle_t$).}
    \label{fig:heatmap_yearlyvariations}
\end{figure}

Figure \ref{fig:heatmap_yearlyvariations} shows the standard variation of the total equivalent annual system costs $Q_{tot}$ between different years for different values of the wind share $\alpha$ and the generation factor $\gamma$. One can see that the variations are lowest with a low wind share $\alpha$=0.5 and low generation factor $\gamma$=0.9. In this case, a significant share of the energy is provided by backup power (cf.\  figure \ref{fig:histogram_cost_basecase}) which allows to overcome variations in the generation and load patterns to a large extent. Increasing the wind share $\alpha$ and the generation factor $\gamma$ increases the variations in the total equivalent annual costs $Q_{tot}$, reaching a standard variation of up to 4\%. In absolute values and considering the variations in the original data sets, this corresponds to e.g.\ total equivalent annual system costs varying between $Q_{tot}\approx$~38.9 bn EUR for the year 2009 to $Q_{tot}\approx$~43.8 bn EUR for the year 2010 in scenarios with a wind share $\alpha$=0.8 and a generation factor $\gamma$=1.0. In comparison to the seven-years annual system costs of $Q_{tot}\approx$~41.1 bn EUR, these variations correspond to a deviation of up to 7\%. In terms of the installed storage capacities, the differences between individual years lead to e.g.\ a discharging power $P_{H2S}^{DL}$ for the hydrogen storage of $P_{H2S}^{DL}\approx$~10.0~GW for the year 2009 and $P_{H2S}^{DL}\approx$~17.0~GW for the year 2010, while a value of $P_{H2S}^{DL}\approx$~12.0~GW was found for the seven-years data in the optimal case. The system component which was found to vary most between different years is the size of the hydrogen storages $H_{H2S}^{\max}$ with $H_{H2S}^{\max}\approx$~119~av.l.h.\ for the year 2010 and $H_{H2S}^{\max}\approx$~364~av.l.h.\ for the year 2006. It is worth noting that even the smallest value was found to be  higher than the value for the seven-years data $H_{H2S}^{\max}\approx$~113~av.l.h.\, indicating that particularly the required size of the hydrogen storage is overestimated when individual years are considered in comparison to long-term data.

Overall, these results illustrate the large variations in the underlying generation patterns for wind and solar energy and consequently the need to consider multi-year data when studying prospective power supply systems. Furthermore, this result implies that the previously found low average full-load hours might be even lower in some years, so that some backup power capacities are not used for a year or longer while always being available on short-term basis.

%%%%%%%%%%%%%%%%%%%%%%%%%%%%%%%%%%%%%%%%%%%%%%%%%%%%%%%%%%%%%%%%%
%% CONCLUSION
%%%%%%%%%%%%%%%%%%%%%%%%%%%%%%%%%%%%%%%%%%%%%%%%%%%%%%%%%%%%%%%%%

\section{Conclusions}

Future scenarios based on RES will have high shares of electricity originating from variable resources like wind and solar energy. To guarantee a stable system, measures have to be taken to balance load and generation at all times. In this work, we looked at prospective power supply systems with very high shares of VRES, in which the fluctuations are balanced using a combination of storage devices as well as fully-flexible backup power plants and curtailment. We presented a modelling approach to systematically study which combination of these balancing options leads to minimum costs. 

In accordance with results from related works, we found that the highest shares in costs are related to the generation capacities, while only a minor share is related to the storage costs. For the storage, the costs for capacities were found to be generally lower compared to the costs for charging and discharging. Furthermore, our results indicate that storage based on synthetic hydrogen as well as pumped-hydro storages might be favourable to be installed in scenarios in which a significant share of the VRES generation originates from wind. Lithium-ion storage with its high efficiencies but high capacity costs was only found to be present in case of scenarios with high shares of PV. In this vein, we can confirm that the daily fluctuations in the solar scenarios lead to a need for efficient, small daily storage, whereas seasonal storage are required for scenarios dominated by wind energy. This implies that the cheap capacity costs of hydrogen storage can partly overcome its low efficiency. Investigating scenarios without hydrogen storage being available significantly increased the overall system costs, which underlines the importance of this technology.

We also took a closer look at backup power need and found that backup would be needed with high backup power up to about 30 GW. These power plants would only run on average about 2000-3000 equivalent full-load hours. Furthermore, we found that the yearly variations reach values up to 4\%, which corresponds to a difference up to five billion EUR equivalent annual costs between cheapest and most expensive year. In practical terms, these results imply that the system (operator and/or society) has to come up with solutions how to finance those large capacities which are used at rare intervals. Solutions such as demand-side management might help to shift consumption by a few hours, yet concepts need to be found how to deal with long periods of dead calm which occur only once every years and how many backup capacities should be kept as final reserve for this purpose.

We conclude this work with recommendations regarding future work: As we have seen in this work, the results vary significantly depending on the weather situation in the investigated scenario. Hence, future work should also be done with multi-year data so as to include the statistical variations of these long-term weather patterns. Furthermore, as also done in this work, we recommend to not explicitly include the generation costs in the optimisation part but instead study scenarios with different wind-solar-mixes and draw comparisons between them. Regarding the modelling approach presented in this work, a further development could trigger the modelling of the backup process by using e.g.\ an iterative approach to match the price of backup with resulting full-load hours. Including large capacities of controllable power plants like biomass or hydro power as virtual storage might lead to lower storage needs. This way, a stable and affordable power supply system based mainly on Renewable Energy Sources will likely be found.

%%%%%%%%%%%%%%%%%%%%%%%%%%%%%%%%%%%%%%%%%%%%%%%%%%%%%%%%%%%%%%%%%
%% ACKNOWLEDGEMENTS
%%%%%%%%%%%%%%%%%%%%%%%%%%%%%%%%%%%%%%%%%%%%%%%%%%%%%%%%%%%%%%%%%

\section*{Acknowledgements}
The authors thank Thomas Vogt and Lukas Wienholt for helpful discussions. Financial support of the first author through a PhD scholarship by the Lower Saxony Ministry for Science and Culture (MWK) is acknowledged. The authors thank Lueder von Bremen and Alexander Kies (ForWind Centre for Wind Energy Research, University of Oldenburg, Germany) for providing the wind and solar power generation data set used in this work. This data set was developed in the research project RESTORE 2050 (No.\ 03SF0439) funded by the German Federal Ministry of Education and Research in the Energy Storage Funding Initiative.

%%%%%%%%%%%%%%%%%%%%%%%%%%%%%%%%%%%%%%%%%%%%%%%%%%%%%%%%%%%%%%%%%
%% APPENDIX
%%%%%%%%%%%%%%%%%%%%%%%%%%%%%%%%%%%%%%%%%%%%%%%%%%%%%%%%%%%%%%%%%

\appendix
\section{Conversion to equivalent annual costs}\label{sec:app_eac}
\noindent The parameters $q^X$ corresponding to the equivalent annual costs for technology $X$ are composed of $q^{I,X}$, an annual share of the installation costs $I^X$, and the annual operating \& maintenance costs $q^{O,X}$:

\[ q^X=q^{I,X}+q^{O,X} \]

with

\[ q^{I,X}=\frac{I^X}{A_{t,r}}~~\text{where}~~ A_{t,r}=\frac{1-\frac{1}{(1+r)^n}}{r}\,, \]

and $n$ being the lifetime in years and $r$ being the interest rate. An interest rate of $r=0.06$ is used throughout this work. The annual operating \& maintenance costs $q^{O,X}$ are directly related to the overall installation costs through the factor $f^{O,X}$:

\[ q^{O,X}=f^{O,X}\cdot I^{X}\,.\]

The values for $I^X$, $f^{O,X}$ and $n$ are given in table \ref{tab:parameters}.

%%%%%%%%%%%%%%%%%%%%%%%%%%%%%%%%%%%%%%%%%%%%%%%%%%%%%%%%%%%%%%%%%
%% REFERENCES, ETC.
%%%%%%%%%%%%%%%%%%%%%%%%%%%%%%%%%%%%%%%%%%%%%%%%%%%%%%%%%%%%%%%%%

\section*{References}
%\bibliographystyle{unsrt}
%\bibliography{ref.bib}

\begin{thebibliography}{10}

\bibitem{Heide2010}
Dominik Heide, Lueder von Bremen, Martin Greiner, Clemens Hoffmann, Markus
  Speckmann, and Stefan Bofinger.
\newblock Seasonal optimal mix of wind and solar power in a future, highly
  renewable {E}urope.
\newblock {\em Renewable Energy}, 35(11):2483--2489, 2010.

\bibitem{Foundation2010}
European~Climate Foundation.
\newblock Roadmap 2050: A practical guide to a prosperous, low-carbon {Europe}.
\newblock Technical Analysis, 2010.

\bibitem{Hart2011}
Elaine~K Hart and Mark~Z Jacobson.
\newblock A monte carlo approach to generator portfolio planning and carbon
  emissions assessments of systems with large penetrations of variable
  renewables.
\newblock {\em Renewable Energy}, 36(8):2278--2286, 2011.

\bibitem{Budischak2012}
Cory Budischak, DeAnna Sewell, Heather Thomson, Leon Mach, Dana~E Veron, and
  Willett Kempton.
\newblock Cost-minimized combinations of wind power, solar power and
  electrochemical storage, powering the grid up to 99.9\% of the time.
\newblock {\em Journal of Power Sources}, 2012.

\bibitem{Becker2015}
Sarah Becker, Bethany~A. Frew, Gorm~B. Andresen, Mark~Z. Jacobson, Stefan
  Schramm, and Martin Greiner.
\newblock Renewable build-up pathways for the {US}: Generation costs are not
  system costs.
\newblock {\em Energy}, 81:437--445, 2015.

\bibitem{Esteban2012}
Miguel Esteban, Qi~Zhang, and Agya Utama.
\newblock Estimation of the energy storage requirement of a future 100\%
  renewable energy system in {J}apan.
\newblock {\em Energy Policy}, 47:22--31, 2012.

\bibitem{Mason2010}
Ian~George Mason, SC~Page, and AG~Williamson.
\newblock A 100\% renewable electricity generation system for {N}ew {Z}ealand
  utilising hydro, wind, geothermal and biomass resources.
\newblock {\em Energy Policy}, 38(8):3973--3984, 2010.

\bibitem{Connolly2011}
David Connolly, Henrik Lund, Brian~Vad Mathiesen, and M~Leahy.
\newblock The first step towards a 100\% renewable energy-system for {I}reland.
\newblock {\em Applied Energy}, 88(2):502--507, 2011.

\bibitem{DeVries2007}
Bert~JM De~Vries, Detlef~P van Vuuren, and Monique~M Hoogwijk.
\newblock Renewable energy sources: Their global potential for the first-half
  of the 21st century at a global level: An integrated approach.
\newblock {\em Energy Policy}, 35(4):2590--2610, 2007.

\bibitem{Jacobson2011}
Mark~Z Jacobson and Mark~A Delucchi.
\newblock Providing all global energy with wind, water, and solar power, part
  {I}: Technologies, energy resources, quantities and areas of infrastructure,
  and materials.
\newblock {\em Energy Policy}, 39(3):1154--1169, 2011.

\bibitem{Rasmussen2012}
Morten~Grud Rasmussen, Gorm~Bruun Andresen, and Martin Greiner.
\newblock Storage and balancing synergies in a fully or highly renewable
  pan-{E}uropean power system.
\newblock {\em Energy Policy}, 51:642--651, 2012.

\bibitem{Agora2014}
Danel F{\"u}rstenwerth, Lars Waldmann, Michael Sterner, Martin Thema, Fabian
  Eckert, Albert Moser, Andreas Sch{\"a}fer, Tim Drees, Christian Rehtanz, Ulf
  H{\"a}ger, Jan Kays, Andre Seack, Dirk~Uwe Sauer, Matthias Leuthold, and
  Philipp St{\"o}cker.
\newblock Stromspeicher in der {E}nergiewende.
\newblock {Agora Energiewende}, 2012.

\bibitem{Lund2009}
Henrik Lund and Brian~Vad Mathiesen.
\newblock Energy system analysis of 100\% renewable energy systems -- the case
  of {D}enmark in years 2030 and 2050.
\newblock {\em Energy}, 34(5):524--531, 2009.

\bibitem{Schill2013}
Wolf-Dieter Schill.
\newblock Residual load, renewable surplus generation and storage requirements
  in {G}ermany.
\newblock DIW - Deutsches Institut f{\"u}r Wirtschaftsforschung, 2013.

\bibitem{Bremen2009}
Lueder~von Bremen, K~Knorr, B~Lange, and S~Bofinger.
\newblock A fully renewable power supply scenario for {E}urope: The weather
  determines storage and transport.
\newblock In {\em 8th International Workshop on Large Scale Integration of Wind
  Power into Power Systems as well as on Transmission Networks for Offshore
  Wind Farms, Bremen}, 2009.

\bibitem{Heide2011}
Dominik Heide, Martin Greiner, Lueder von Bremen, and Clemens Hoffmann.
\newblock Reduced storage and balancing needs in a fully renewable {E}uropean
  power system with excess wind and solar power generation.
\newblock {\em Renewable Energy}, 36(9):2515--2523, 2011.

\bibitem{Rodriguez2014}
Rolando~A. Rodriguez, Sarah Becker, Gorm~B. Andresen, Dominik Heide, and Martin
  Greiner.
\newblock Transmission needs across a fully renewable {E}uropean power system.
\newblock {\em Renewable Energy}, 63:467--476, 2014.

\bibitem{Weiss2013a}
Thomas Weiss and Detlef Schulz.
\newblock Development of fluctuating renewable energy sources and its influence
  on the future energy storage needs of selected {E}uropean countries.
\newblock In {\em Proceeding of the 4th International Youth Conference on
  Energy (IYCE), 2013}, pages 1--5, June 2013.

\bibitem{Bussar2016}
Christian Bussar, Philipp St{\"o}cker, Zhuang Cai, Luiz~Moraes Jr., Dirk
  Magnor, Pablo Wiernes, Niklas van Bracht, Albert Moser, and Dirk~Uwe Sauer.
\newblock Large-scale integration of renewable energies and impact on storage
  demand in a {E}uropean renewable power system of 2050 - sensitivity study.
\newblock {\em Journal of Energy Storage}, 6:1--10, 2016.

\bibitem{Weitemeyer2014}
Stefan Weitemeyer, David Kleinhans, Thomas Vogt, and Carsten Agert.
\newblock Integration of renewable energy sources in future power systems: The
  role of storage.
\newblock {\em Renewable Energy}, 75:14--20, 2015.

\bibitem{Weitemeyer2016}
Stefan Weitemeyer, David Kleinhans, Lukas Wienholt, Thomas Vogt, and Carsten
  Agert.
\newblock A {E}uropean perspective: Potential of grid and storage for balancing
  renewable power systems.
\newblock {\em Energy Technology}, 4(1):114--122, 2016.

\bibitem{nocedal2006numerical}
Jorge Nocedal and Stephen Wright.
\newblock {\em Numerical optimization}.
\newblock Springer Science \& Business Media, 2006.

\bibitem{grossmann1983effective}
Ch~Grossmann.
\newblock An effective method for solving linear programming problems with
  flexible constraints.
\newblock {\em Zeitschrift f{\"u}r Operations Research}, 27(1):107--122, 1983.

\bibitem{BMWi2016}
{Bundesministerium f\"ur Wirtschaft und Energie (BMWi)}.
\newblock {Erneuerbare Energien in Deutschland -- Daten zur Entwicklung im Jahr
  2015}.
\newblock Brochure, 2016.

\bibitem{Kies2015}
Alexander Kies, Kabitri Nag, Lueder von Bremen, Elke Lorenz, and Deltev
  Heinemann.
\newblock Investigation of balancing effects in long term renewable energy
  feed-in with respect to the transmission grid.
\newblock {\em Advances in Science and Research}, 12(1):91--95, 2015.

\bibitem{Adamek2012}
Franziska Adamek, T~Aundrup, W~Glaunsinger, M~Kleimaier, H~Landinger,
  M~Leuthold, B~Lunz, A~Moser, C~Pape, H~Pluntke, N~Rotering, D~U Sauer,
  M~Sterner, and W~Well{\ss}ow.
\newblock {VDE-Studie} : {E}nergiespeicher f{\"u}r die {E}nergiewende --
  {G}esamttext.
\newblock {\em VDE}, 2012.

\bibitem{Eurelectric2011}
{Eurelectric - Union of the Electricity Industry}.
\newblock Hydro in {Europe}: Repowering renewables.
\newblock Depot Legal: D/2011/12.105/41, 2011.

\bibitem{Steffen2012}
Bjarne Steffen.
\newblock Prospects for pumped-hydro storage in {G}ermany.
\newblock {\em Energy Policy}, 45:420--429, 2012.

\bibitem{Nitsch2010}
Joachim Nitsch, Thomas Pregger, Yvonne Scholz, Michael Sterner, Norman
  Gerhardt, Amany von Oehsen, Carsten Pape, Yves-Marie Saint-Drenan, and Bernd
  Wenzel.
\newblock Langfristszenarien und {S}trategien f{\"u}r den {A}usbau der
  erneuerbaren {E}nergien in {D}eutschland bei {B}er{\"u}cksichtigung der
  {E}ntwicklung in {E}uropa und global - {L}eitstudie 2010.
\newblock Arbeitsgemeinschaft DLR, IWES, IFNE, 2010.

\bibitem{Nitsch2012c}
Joachim Nitsch, Thomas Pregger, Yvonne Scholz, Michael Sterner, Norman
  Gerhardt, Amany von Oehsen, Carsten Pape, Yves-Marie Saint-Drenan, and Bernd
  Wenzel.
\newblock Langfristszenarien und {S}trategien f{\"u}r den {A}usbau der
  erneuerbaren {E}nergien in {D}eutschland bei {B}er{\"u}cksichtigung der
  {E}ntwicklung in {E}uropa und global - {D}atenanhang {II} zum
  {S}chlussbericht.
\newblock Arbeitsgemeinschaft DLR, IWES, IFNE, 2012.

\bibitem{Kost2013}
Christoph Kost, Johannes~N. Mayer, Jessica Thomsen, Niklas Hartmann, Charlotte
  Senkpiel, Simon Philipps, Sebastian Nold, Simon Lude, and Thomas Schlegl.
\newblock Stromgestehungskosten {E}rneuerbare {E}nergien.
\newblock Fraunhofer-Institut f\"ur solare {E}nergiesysteme ISE, 2013.

\end{thebibliography}

\end{document}